# Influence of Mg Deficiency on the Superconductivity in MgB$_2$ Thin Films Grown by using HPCVD

**Ji Young Huh, Won Kyung Seong, Soon-Gil Jung, and W. N. Kang**
*BK21 Division and Department of Physics, Sungkyunkwan University, Suwon 440-746, Korea*

The effects of Mg deficiency in MgB$_2$ films grown by using hybrid physical-chemical vapor deposition were investigated after vacuum annealing at various temperatures. High-quality MgB$_2$ films grown on *c*-cut Al$_2$O$_3$ substrates with different superconducting transition temperatures (T$_c$) of 40.2 and 41 K were used in this study. As the annealing temperature was increased from 200 to 800 ℃, the Mg contents in the MgB$_2$ films systemically decreased, but the T$_c$'s did not change, within ± 0.12 K, until the annealing temperature reached 700 ℃. For MgB$_2$ films annealed at 800 ℃ for 30 min., however, no superconductivity was observed, and the temperature dependence of the resistivity showed a semiconducting behavior. We also found that the residual resistivity ratio decreased with increasing annealing temperature.

(Some figures in this article are in colour only in the electronic version)

## 1. Introduction

Since the discovery [1] of superconductivity in intermetallic MgB$_2$ compounds, a number of groups have attempted to enhance the superconducting transition temperature (T$_c$) by electron doping [2, 3] with chemical substitution of Al and C for Mg and B on MgB$_2$, respectively, and by hole doping [3, 4] with substitution of Li for Mg. However, those efforts were not successful in increasing the T$_c$ of MgB$_2$ superconductors. In the meantime, enhancement of T$_c$ (40.3 K) in MgB$_2$ bulks from the $^{10}$B isotope effect was reported by Hinks *et al*. [5], and a higher T$_c$ (41.7 K) in MgB$_2$ thin films fabricated by exposing boron crystals to magnesium vapor was reported by Hur *et al*. [6]. Recently, Pogrebnyakov *et al*. [7] reported a thickness dependence of T$_c$ in MgB$_2$ thin films grown on SiC and Al$_2$O$_3$ substrates by using hybrid physical-chemical vapor deposition (HPCVD) and observed the highest values of T$_c$, 41.8 K and 40.5 K, for MgB$_2$ films deposited on SiC and Al$_2$O$_3$ substrates, respectively. They proposed that the enhancement of T$_c$ in MgB$_2$ films was due to a softening of a bond-stretching phonon mode with a biaxial tensile strain.

Among the numerous growth techniques of MgB$_2$ thin films, such as HPCVD [8], pulsed laser deposition (PLD) [9], reactive evaporation [10], and molecular beam epitaxy [11], only the films fabricated by using HPCVD showed high-T$_c$ well above the bulk value [7]. Our MgB$_2$ thin films of 1 µm in thickness deposited on Al$_2$O$_3$ also showed a T$_c$ of 41 K, consistent with the general trend of the thickness dependence of T$_c$ reported by Pogrebnyakov *et al*. [7]. Moreover, the residual resistivity ratio (RRR) in MgB$_2$ compounds was observed to have various values [8-11] as high as 30, depending on the preparation conditions, and the larger RRR value, the higher T$_c$, suggesting that Mg concentrations could have an effect on T$_c$. Even a RRR value of 80 with a residual resistivity of 0.1 µΩ cm in an MgB$_2$ film was recently reported [12]. Therefore, one may expect the Mg deficiency dependence of T$_c$ and the RRR values for higher T$_c$ thin films grown by using HPCVD to provide very crucial results in understanding the above issues.

In this paper, we report the effects of Mg deficiency in MgB$_2$ films grown by using HPCVD, and vacuum-annealed at various temperatures. We found that the T$_c$'s did not change with increasing annealing temperature up to 700 ℃ and that no superconductivity was observed after annealing at 800 ℃. The RRR values decreased slightly with increasing annealing temperature up to 500 ℃ and then abruptly decreased at higher annealing temperatures above 600 ℃.

## 2. Experimental

The HPCVD system used in this study has been described in detail elsewhere [13]. We have chosen two types of thin films grown on (0001) Al$_2$O$_3$ with different thicknesses of 1.0 μm (M1.0um) and 2.2 μm (M2.2um) and different onset transition temperatures ($T_{c,on}$) of 41 K and 40.2 K, respectively. These two samples were deposited under the same conditions except that the growth temperatures of 630 and 580 ℃ for the M1.0um and the M2.2um samples, respectively. By using an induction heater, we evaporated Mg chips of 2 – 3 mm in diameter at the growth temperatures and we introduced 5% $B_2H_6$ in $H_2$ at a flow rate of 50 sccm as the B precursor. The carrier gas was $H_2$ at a flow rate of 100 sccm, and the total pressure in the growth chamber was kept at 150 Torr.

X-ray diffraction (XRD) patterns indicated c-axis-oriented epitaxial growth for those two films. The typical sample size used in the present study was 5 mm x 2 mm. Since we annealed one sample several times at different temperatures and measured the temperature dependence of the resistance and the composition of $MgB_2$ by using energy dispersive spectroscopy (EDS) after each annealing, it was really important not to change any part of the measurement configuration, such as the silver pads and the locations of the contact wires. In most cases, changes in the measurement configuration gave different transport properties. Careful preparation of the electrical contacts by using silver coating under high vacuum (< $10^{-6}$ Torr) was crucial for obtaining a low contact resistance. In order to reduce systematically the Mg concentration in pristine samples, we carried out high-vacuum (~ $10^{-6}$ Torr) annealing at temperatures from 200 °C to 800 °C in 30 minutes. After each annealing, we measured the temperature dependence of the resistance by using the dc four-probe method at an applied current of 10 mA.

**3. Results and discussion**

The typical temperature dependence of the resistivity curves normalized to the resistivity at 42 K for the as-grown $MgB_2$ thin films of M1.0um and M2.2um are shown in Fig. 1 (a). The residual resistivity ratio (RRR = $\rho_{300K}/\rho_{42K}$) value (26.4) of M1.0um sample was observed to be about 6 times larger than that (4.1) of the M2.2um sample. The residual resistivities at 42 K were 0.32 and 3.8 μΩcm for the M1.0um and the M2.2um samples, respectively. The inset shows a magnified view near $T_c$. We can see a $T_c$ of 41 for the M1.0um sample, which is 2 K higher than the bulk value and which is the highest superconducting transition temperature of the $MgB_2$ thin films deposited on $Al_2O_3$. The c-axis lattice constant of both films obtained from the XRD data was 3.512 Å, which is smaller than the value (3.52 Å) of bulk $MgB_2$ [1] and close to the value of higher-$T_c$ $MgB_2$ films on SiC reported by Pogrebnyakov *et al.* [7]. For the thicker M2.2um film, however, we observed a $T_c$ of 40.2 K lower than that of the M1.0um sample, which is probably due to the low-temperature growth at 580 °C. After annealing at 800 °C, the temperature dependences of resistivity for both the M1.0um and the M2.2um samples showed semiconducting behaviors, as shown in Fig. 1 (b), and no superconductivity was observed until the temperature had been lowered to 10 K.

Figure 2 shows the temperature dependences of the resistivities before and after annealing at 400, 500, and 600 °C for (a) the M1.0um and (b) the M2.2um thin films. Generally, the resistivity increased with increasing annealing temperatures for both samples, which indicated that Mg diffused out from the $MgB_2$ thin films because of the baking under high vacuum. For the samples annealed below 500 °C, the resistivity of the M1.0um film changed weakly compared to that of the M2.2um film because the M1.0um film had an Mg film (~ 50 nm in thickness) coated on the top of the $MgB_2$ thin film, as shown in the inset of Fig. 2 (b). The large RRR values of the M1.0um films originated partially from the Mg layer on the surface of the thin film [14]. An $MgB_2$ thin film covered by an Mg layer was obtained by using a quenching process within 2 minutes after film deposition. On the other hand, we fabricated pure $MgB_2$ thin films without Mg layers by using a slow cooling process at about 10°C/min. under a pressure of 150 Torr with flowing $H_2$ gas.

Figure 3 shows the relative Mg ratio (%) of the M2.2um film normalized to the quantity of Mg atoms in the as-grown film as a function of annealing temperature from 200 to 800 °C. The inset is an SEM image of the M2.2um film in which the silver pads for the resistivity measurements are denoted by Ag. The Mg concentration was measured by EDS in the rectangular (1.2 x 1.9 $mm^2$) area located between the Ag pads. The EDS analysis was always carried out for the same area after each annealing so that we were able to directly compare the relative Mg ratios by normalizing them to the quantity of Mg atoms in the as-grown film. We can clearly see that the relative Mg ratio linearly decreased to 87% with increasing annealing temperatures up to 700 °C and then abruptly decreased to 66% after annealing at 800 °C.

Figure 4 shows the annealing temperature dependence of (a) the onset transition temperatures ($T_{c,on}$) and (b) the RRR values for the M1.0um and the M2.2um thin films. The RRR data for M2.2um sample were multiplied by 5 for the sake of clear comparison with those of the M1.0um sample. For both samples, the $T_{c,on}$'s were observed to be almost independent of the annealing temperatures up to 700 °C while the RRR values decreased gradually with increasing annealing temperature up to 500 °C and then decreased dramatically after annealing at temperatures above 600 °C. These results indicate that Mg initially diffused out from the surfaces of the $MgB_2$ films and that vacuum annealing at high temperatures above 500 °C could have a critical effect on the superconductivity. Even for the higher-

$T_c$ samples, the Mg deficiency seemed to be less sensitive to $T_c$, consistent with the data reported for polycrystalline bulk samples [15, 16]. Our data suggest that a low-temperature process (below 500 °C) is required in case of fabricating superconducting devices with insulating or metallic layers.

## 4. Summary

By using high-$T_c$ MgB$_2$ thin films with $T_c$ values of 40.2 and 41 K, we have investigated the effect vacuum annealing on the superconductivity and the RRR. We found that the Mg content in the MgB$_2$ films systemically decreased as the annealing temperature increased from 200 to 800℃, but the values of the $T_c$'s did not change until the annealing temperature reached 700℃. The superconductivity was broken after annealing at 800℃ for 30 minutes, and eventually the temperature dependence of resistivity showed a semiconducting behavior. The RRR decreased slightly with increasing annealing temperature up to 500 ℃ and then abruptly decreased for higher annealing temperatures above 600 ℃, suggesting that a low-temperature process is required when fabricating superconducting devices with insulating or metallic layers.

**Acknowledgements**
This work was supported by the Korea Research Foundation grant funded by the Korean Government (MOEHRD) (KRF-2005-005-J11902 & KRF-2006-312-C00130) and by the Korea Science and Engineering Foundation (KOSEF) grant funded by the Korea Government (MOST) (R01-2005-000-11001-0). This research was partially supported by a grant (R-2006-1-248) from Electric Power Industry Technology Evaluation & Planning (ETEP), Republic of Korea.

**Figure and Figure captions**

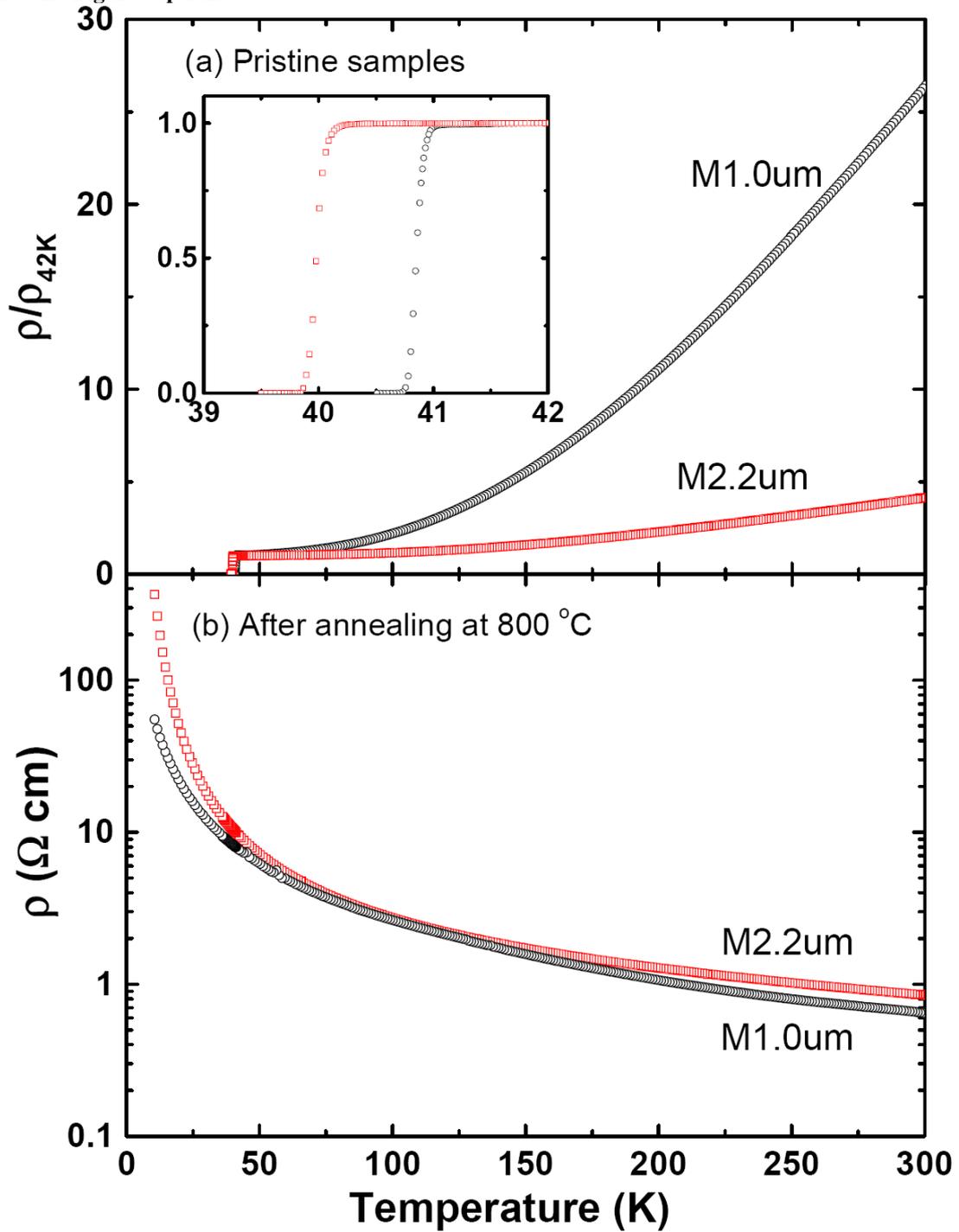

Fig. 1. Temperature dependence of resistivity curves around $T_c$ for $MgB_2$ thin films (a) before and (b) after vacuum annealing at 800 °C. The inset shows a magnified view near $T_c$ for sake of clarity.

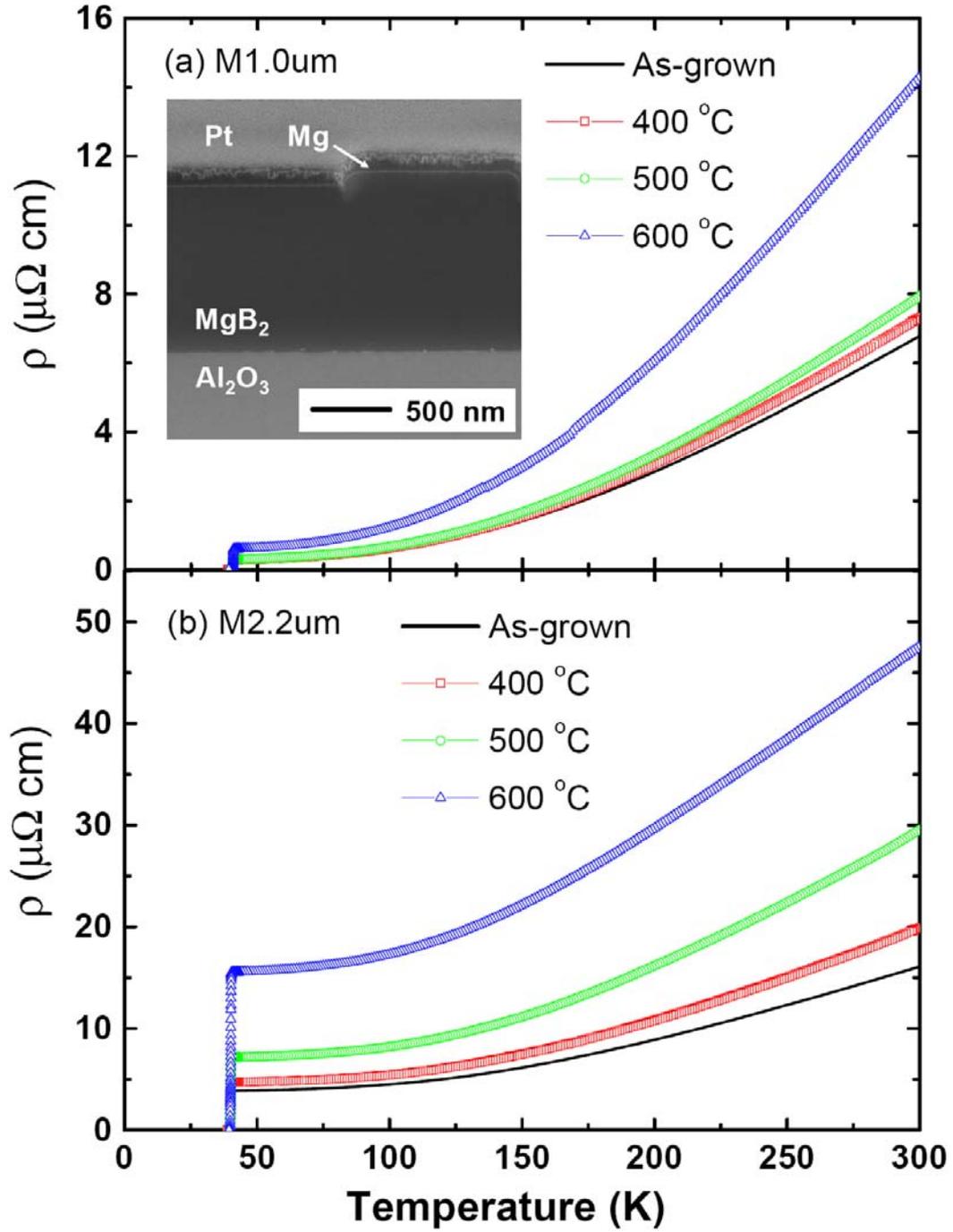

Fig. 2. Resistivity vs temperature curves of (a) the M1.0um and (b) the M2.2um films for as-grown films (solid lines) and for films vacuum annealed at 400 (squares), 500 (circles), and 600 °C (triangles). The inset shows a cross-sectional view of an SEM image for the M1.0um film, where the excess Mg layer can be clearly seen on the surface of the $MgB_2$ film.

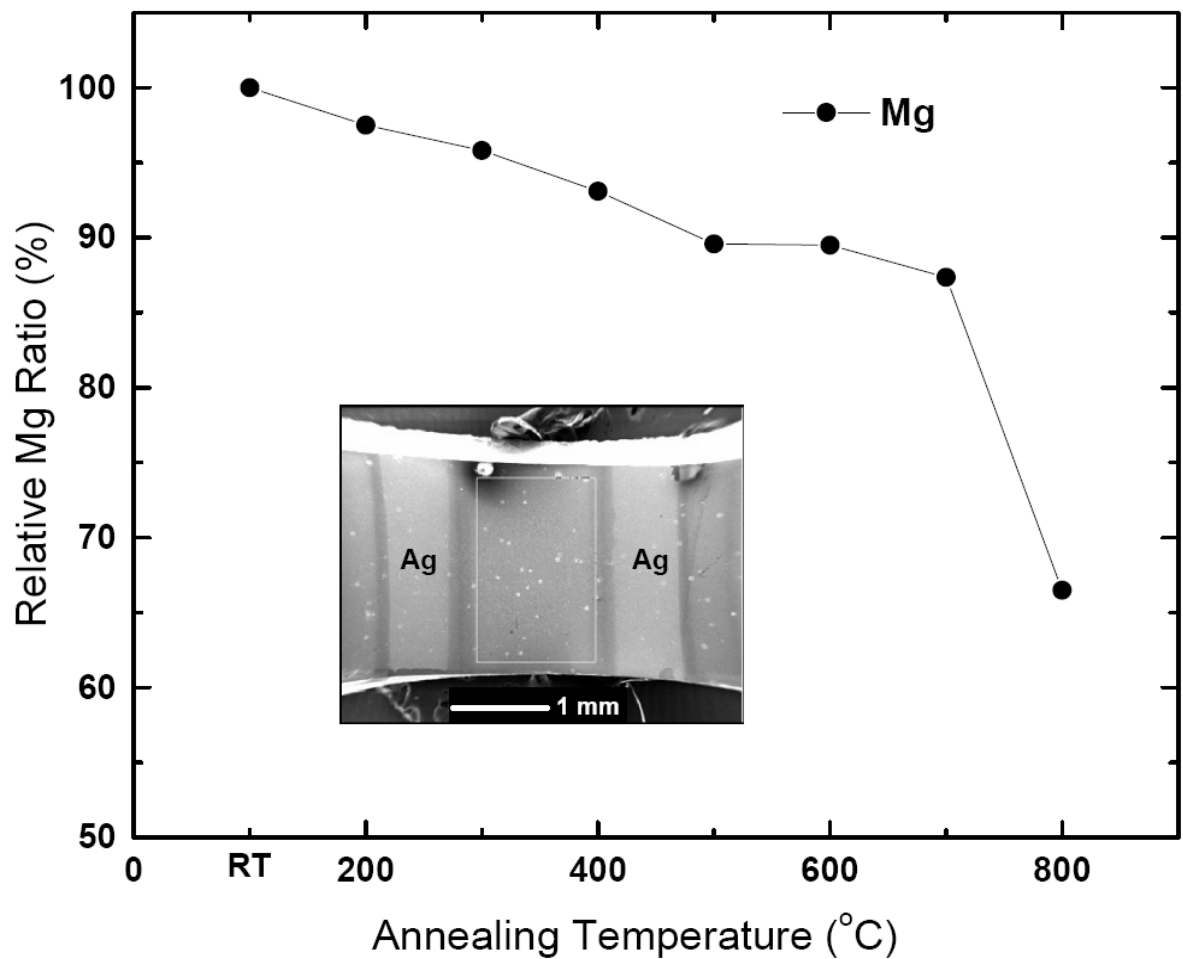

Fig. 3. Vacuum-annealing temperature dependence of the relative Mg ratio normalized to the quantity of Mg atoms in the as-grown film. The inset is an SEM image of the M2.2um film in which silver pads are denoted by Ag. The Mg concentration was measured by using EDS in the rectangular (1.2 x 1.9 mm$^2$) area located between Ag pads.

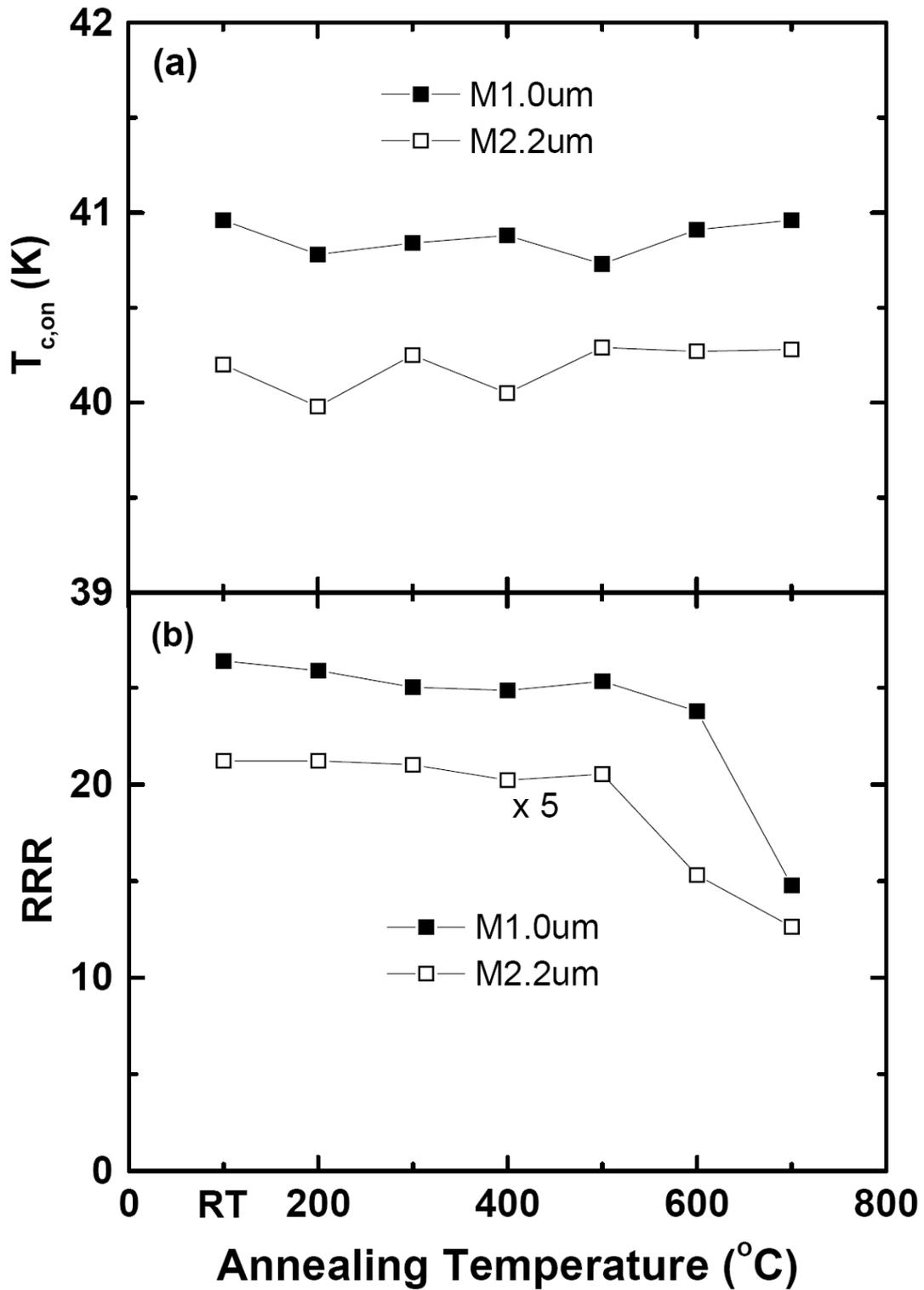

Fig. 4. Vacuum-annealing temperature dependence of (a) $T_{c,on}$ and (b) RRR for the M1.0um and the M2.2um samples. The RRR values for the M2.2um sample were multiplied by 5 for the sake of clarity.